\documentclass[aps]{revtex4}
\usepackage{graphicx,bm,psfrag}
\begin{document}

\title{ Refraction of a Gaussian Seaway}
\author{E. J. Heller$^{1,2}$, L. Kaplan$^{3}$, and A. Dahlen$^{1}$}
\affiliation{$^1$ Department of Physics, Harvard University, Cambridge, MA 02138, USA \\
$^2$ Department of Chemistry and Chemical Biology, Harvard University, Cambridge, MA 02138, USA \\
$^3$ Department of Physics, Tulane University, New Orleans,
LA 70118, USA}

\date{December 28, 2007}

\begin{abstract}
Refraction of a Longuet-Higgins Gaussian sea by random ocean currents creates persistent local variations in average energy and wave action. These variations take the form of lumps or streaks, and they explicitly survive dispersion over wavelength and incoming wave propagation direction. Thus, the uniform sampling assumed in the venerable Longuet-Higgins theory does not apply following refraction by random currents. Proper handling of the non-uniform sampling results in greatly increased probability of freak wave formation. The present theory represents a synthesis of Longuet-Higgins Gaussian seas and the refraction model of White and Fornberg, which considered the effect of currents on a plane wave incident seaway. Using the linearized equations for deep ocean waves, we obtain quantitative predictions for the increased probability of freak wave formation when the refractive effects are taken into account. The crest height or wave height distribution depends primarily on the ``freak index", $\gamma$, which measures the strength of refraction relative to the angular spread of the incoming sea. Dramatic effects are obtained in the tail of this distribution even for the modest values of the freak index that are expected to occur commonly in nature. Extensive comparisons are made between the analytical description and numerical simulations.
\end{abstract}

\maketitle

\section{Introduction}
\label{secintro}

Freak waves in the ocean, known also as rogue waves or giant waves, are waves of extreme height relative to the typical wave in a given sea state, and may arise during a storm or in relatively calm seas. Due to their steepness, such waves pose great risk to cargo ships and even to large cruise liners. Well-publicized and documented encounters include a 25.6 meter wave that hit the Draupner oil platform in the North Sea in 1995, two ships that suffered damage at 30 meters above sea level from a single wave in the South Atlantic in 2001, and the cruise liner Norwegian Dawn that met a series of three 21 meter waves off the coast of Georgia in 2005. In March 2007, the MS Prinsendam was hit by a 21 meter tall freak wave in the Antarctic. Satellite images taken over three weeks in 2001 and analyzed as part of the European Union MaxWave project~\cite{maxwave} detected ten waves of height above 25 meters, suggesting that such waves commonly occur in the world's oceans.

Random (constructive) linear superposition of many plane waves with differing direction and wavelength, as in the Longuet-Higgins random seas model~\cite{randseas}, offers a simple statistical explanation for the occurrence of freak waves. By the central limit theorem, the sea surface height in this model must be a Gaussian random variable with some standard deviation $\sigma$. In the limit of a narrow frequency spectrum the crest height then follows a Rayleigh distribution: the probability of crest height exceeding $H$ is given by
\begin{equation}
\label{rayleigh}
P_{\rm Rayleigh}(H) = e^{-H^2/2 \sigma^2} \,.
\end{equation}
Note that for linear waves there is an exact symmetry between crests and troughs, so that for a crest height of $H$, the wave height (crest to trough) is given by $2H$. Conventionally, a freak wave is defined as $H \ge 4.4 \sigma$, or $2H \ge 2.2\,{\rm SWH}$, where the significant wave height ${\rm SWH} \approx 4.0 \sigma$ is the average of the largest one third of wave heights in a time series~\cite{swh}. The Rayleigh distribution thus predicts freak waves to occur with probability $6.3 \cdot 10^{-5}$ and extreme freak waves of crest height $H \ge 6\sigma$ (or $2H \ge 3\,{\rm SWH}$) to occur with probability $1.5 \cdot 10^{-8}$. Observational data~\cite{maxwave} suggests that this purely stochastic Rayleigh model significantly underestimates the actual number of freak waves. A review of several alternative theories of the freak wave phenomenon appears in Ref.~\cite{kharif}.

Nonlinear instability effects~\cite{nonlin1,nonlin2} have been extensively and successfully studied as a mechanism of freak wave formation. However, the effects of such instabilities depend sensitively on initial conditions, and the full numerical computations starting from a generic random sea state are costly. Thus, it is difficult to obtain quantitative predictions of the crest height distribution, analogous to (\ref{rayleigh}), except in approximations such as the Nonlinear Schr\"odinger Equation (NLS) or the Dysthe equation~\cite{nonlin2,dysthe79}, which are valid for small to moderate values of the wave steepness. Since nonlinear effects scale as powers of the wave steepness $kH$, where $k$ is the wave number, strongly nonlinear evolution is more likely to be triggered in an initial condition where the waves are already unusually high. Thus the tail of the crest height distribution may be dominated by linear triggering mechanisms that produce slightly taller-than-average waves especially conducive to nonlinear instability. Identifying such triggering mechanisms, and combining them with nonlinear evolution, would allow for quantitative predictions of the freak wave probability distribution, and would also advance the long-term goal of freak wave forecasting.

One such linear triggering mechanism for deep water waves is the focusing or refraction of an incoming plane wave by random current eddies, as discussed by several authors including Peregrine~\cite{peregrine} and White and Fornberg~\cite{wf}, and motivated by the fact that many freak waves have been observed in regions of strong ocean currents. This mechanism must contain part of the story, since ocean currents do refract waves in the ocean. The difficulty is with the assumption of a single plane wave incident on the refracting currents. A plane wave leads to caustics or singularities with infinite ray density (smoothed out only at the wavelength scale), and consequently a repeated and reproducible pattern of freak waves, which will {\it always} appear whenever focusing currents are present. Statistical predictions have no place in such a theory. Of course, an incoming plane wave is a physically unrealistic model of most sea states, since directional and wave number spread in the incoming sea will smear out the above-mentioned singularities.  

These difficulties may be overcome by combining the stochastic random seas approach and the focusing approach, to obtain the distribution of crest heights resulting from a {\it random} incoming sea incident on a region of random eddy currents. Averaging over a random incoming sea smears out but does not entirely destroy the effects of the refraction.  Lumps in wave action persist, varying from place to place by factors of 2 to 4 typically. For realistic values of the parameters, the predicted probability distribution of crest heights depends on a single quantity, the ``freak index," which is a simple function of mean wave speed, rms current speed, and angular spread of the incoming waves.  We shall see that realistic sea states can lead to enhancements of a factor of 50 or more in the probability of forming freak waves, over and above purely Gaussian seas with the same averaged action density.  The greatest probability enhancement occurs for the largest freak waves, while the probabilities associated with typical wave heights are scarcely affected. 

In Sec.~\ref{secthy} we review the focusing model, combine it with a stochastic incoming sea, and obtain analytical results for the probability distribution of crest heights when the freak index is small. The predictions are tested in Sec.~\ref{secnumer} using numerical simulations in the ray limit (where the wavelength is small compared to the size of the eddies). The quantitative correspondence between ray dynamics and wave dynamics is discussed further in Sec.~\ref{secschr}, in connection with analogous correspondence for the Schr\"odinger equation. We emphasize that the probability distributions obtained in Sec.~\ref{secthy} and verified numerically in Sec.~\ref{secnumer} are not the final word, but must rather be used as input to the full nonlinear evolution. This and other outstanding questions are explored in Sec.~\ref{secconc}.

\section{Theory}
\label{secthy}

\subsection{Refraction}
\label{secrefr}
We begin by briefly reviewing the focusing of a plane wave by a random current field. A detailed discussion may be found in the work of White and Fornberg~\cite{wf}.

The ray dynamics of deep-water surface gravity waves is governed by the usual eikonal equations for ray position $\vec r$ and wave vector $\vec k$,
\begin{equation}
{d\vec k\over dt} = -{\partial \omega\over \partial \vec r}; \ \ \ {d\vec r\over dt} = {\partial \omega\over \partial \vec k} \,.
\label{eikonal}
\end{equation}
Here the dispersion relation is given by
\begin{equation}
\omega(\vec k,\vec r) = \sqrt{g\vert \vec k\vert} + \vec k \cdot \vec U(\vec r) \,,
\label{dispers}
\end{equation}
where $\vec U(\vec r)$ is the time-independent current velocity, assumed to be slowly varying on the scale of a wavelength. 
In the following analysis, we consider $\vec U(\vec r)$ to be a random field, with zero average velocity, typical velocity fluctuations of size $u_0$, and spatial correlations on a distance scale $\xi$. In the ocean, the typical eddy size $\xi$ may vary between $20$ and $100$ km, while the typical speed $u_0$ is generally below $1$ m/s.

An incoming plane wave of frequency $\omega$ moving in the $y$ direction may be represented in phase space by initial conditions $\vec r=(x,0)$ for all $x$ and $\vec k =(0,k)$, where $k=2\pi/\lambda$ is related as usual to the wave speed by $v={\partial \omega}/{\partial k}=(g/4k)^{1/2}$ and to the wave period as $T=2\pi/\omega=2\pi(gk)^{1/2}$ (here we neglect the current velocity $\vec U$, which generally is much less than $v$, but the exact expression (\ref{dispers}) is used to relate $\omega$, $v$, and $k$ in numerical simulations, causing $v$ and $k$ to be spatially varying even for a constant frequency $\omega$). A typical wave period for deep-water ocean waves is $10$ s, corresponding to wavelength $\lambda=156$ m and wave speed $v=7.81$ m/s.

\begin{figure}[ht]
\centerline{\includegraphics[width=3.8in,angle=0]{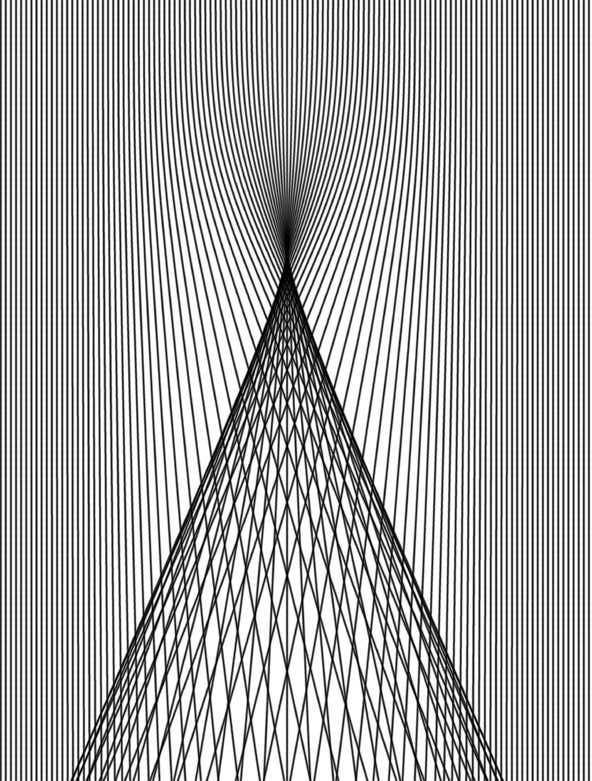}}
\vskip 0.2in
\caption{A cusp singularity, followed by two branches of fold singularities, is formed as initially
parallel rays pass through a focusing region. The two branches appear because the focal
distance varies with the distance of approach from the center, as in a `bad' lens with strong spherical aberration. After
averaging over incident directions, the singularities will
be softened but not washed away completely.}
\label{figcusp}
\end{figure}

When this incoming plane wave impinges on a random current field, it will undergo small-angle scattering, with scattering angle $\sim u_0/v$ after traveling one correlation length $\xi$ in the forward direction. Eventually, singularities appear that are characterized in the surface of section map $[x(0),k_x(0)]\to [x(y),k_x(y)]$ by $\delta x(y) / \delta x(0)=0$, i.e., by local focusing of the manifold of initial conditions at a point. The formation mechanism of these singularities may be ascribed to a `bad lens.' Whereas a good lens without aberration focuses all parallel incoming rays to one point, a bad lens only focuses at each point an infinitesimal neighborhood of nearby rays, so that different neighborhoods get focused at different places as the phase-space manifold evolves forward in $y$, resulting in lines, or branches, of singularities. The typical pattern is an isolated cusp singularity, $\delta^2 x(y) / \delta x(0)^2=0$, followed by two branches of fold singularities, as shown in Fig.~\ref{figcusp}.

The first cusp singularities appear after a median travel distance $y = L \sim \xi (u_0/v)^{-2/3} \gg \xi$ through the random eddy field~\cite{wf,lkbranch}, when the typical ray excursion in the transverse $x$ direction becomes of order $\xi$. For realistic parameters, $L \sim 100$ km or more is typical. [Note that the weakness of the scattering requires a wave to pass over many uncorrelated eddies before the first singularities are formed. This allows ray propagation to be described statistically without regard to detailed structure of individual eddies, using only the length scale $\xi$ and dimensionless velocity ratio $u_0/v$.] The singularities formed in this way are separated by distance $\sim \xi$ in the transverse direction, and the typical deflection angle by the time these singularities appear scales as
\begin{equation}
\delta \theta \approx {\delta k_x \over k} \sim (u_0/v)^{2/3} \,.
\label{delkx}
\end{equation}
The quantity $\delta \theta$ may be defined precisely as the rms value of the deflection angle evaluated at the forward distance $L$, where the averaging is performed over all rays and over an ensemble of random eddy fields. The typical deflection angle $\delta \theta$ does not depend on the size of the eddies but only on the velocity ratio $u_0/v$: faster currents cause larger deflection. For example, for $v=7.81$~m/s and $u_0=0.5$~m/s, we find $(u_0/v)^{2/3}=0.16$, $\delta \theta=18^\circ$, and the median distance to the first singularity is $L=7.5\xi$ ($150$~km if we take the eddy correlation length $\xi$ to be $20$~km).

Of course, the singularities are a mathematical construct existing only in the ray limit $k_x\xi \to \infty$. For wave dynamics, any such singularities must be smeared out at least on the scale of a wavelength, or more precisely, on the scale $k_x^{-1} \sim k^{-1} (u_0/v)^{-2/3}$ in the transverse direction. For example, a fold singularity will be softened over a distance scale $k_x^{-1} (k_x \xi)^{1/3}$~\cite{soften}, resulting in a maximum wave intensity enhanced by a finite factor $(k_x \xi)^{1/3} \sim (k\xi)^{1/3}(u_0/v)^{2/9}$ compared to the background intensity, which is a factor of $5$ or more for reasonable-sized eddies. Although infinities are thus absent from the wave dynamics, this model is still unsatisfactory, as it predicts that extreme freak waves will {\it always} occur whenever an incoming plane wave encounters a random eddy field.

Subsequent propagation through the random current field produces new singularities, and the resulting number of branches grows as $e^{y/L}$ while the wave continues to travel forward. As we will see in Sec.~\ref{secavg} below, wave intensity peaks associated with these later singularities become increasingly washed out due to growing spread in wave direction. Thus, the early generations of caustics, appearing soon after the incoming wave is first scattered by the eddy currents, will typically dominate the freak wave distribution.

\subsection{Averaging}
\label{secavg}

We now consider the more realistic situation where the initial plane wave is replaced by a random superposition of waves with a finite angular spread $\Delta \theta$. Typically, $\Delta \theta$ may take values of $10$ to $30$ degrees. Again, for $k_x \xi \gg 1$, we are justified in modeling the wave dynamics using the ray approximation, where the initial positions are still given by $\vec r=(x,0)$ uniformly distributed over all $x$ and the initial wave vectors are given by $\vec k=(k \sin \theta,k \cos\theta)$, with a Gaussian directional spread $P(\theta) \sim e^{-\theta^2/2(\Delta \theta)^2}$. [In Sec.~\ref{secschr} we confirm the quantitative correspondence between wave and ray intensity statistics in the context of the linear Schr\"odinger equation.]

\begin{figure}[ht]
\centerline{\includegraphics[width=3.8in,angle=0]{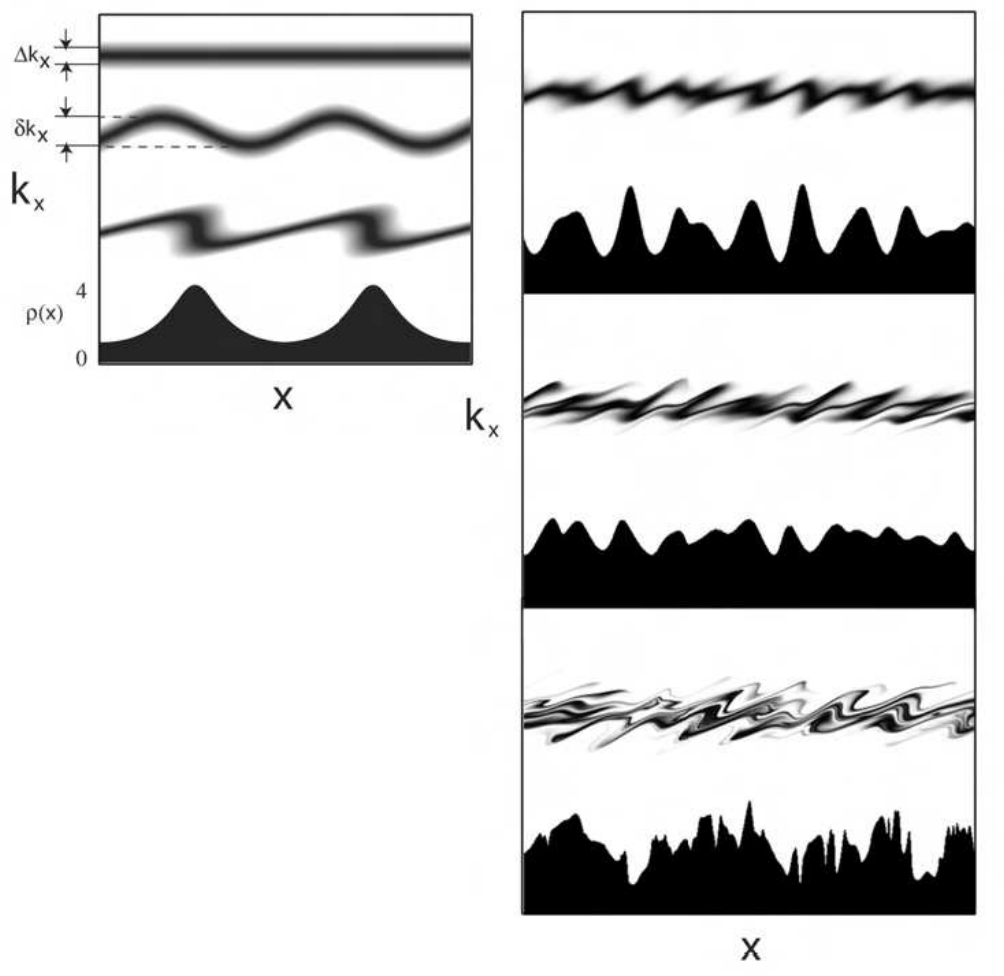}}
\vskip 0.2in
\caption{Left: An initial sea state characterized by a spread $\Delta k_x\approx k \Delta \theta$ in the transverse wave vector is described by a strip of finite thickness in the $(x,k_x)$ phase space at $y=0$ (top). Under subsequent refraction through the eddy field, the sea state acquires additional fluctuations of typical size $\delta k_x \approx k \delta \theta$ as it evolves forward in $y$, and eventually singularities in the ray dynamics are encountered. Although all singularities are washed out due to the finite initial spread $\Delta k_x$, a pattern of high intensity (focusing) regions and low intensity (defocusing) regions is evident when the evolved phase space distribution is projected onto position space. Right: Further refraction through the random eddy field produces multiple tendrils in the phase space distribution and many high and low intensity regions in position space, which may be described statistically.}
\label{figsmooth}
\end{figure}

This set of initial rays is refracted in its evolution through the eddy field, and by the time the first singularities appear, the typical ray is scattered by a transverse wave vector of size $\delta k_x$, as described by (\ref{delkx}). Due to the finite spread $\Delta k_x \approx k \Delta \theta$ of initial conditions, the singularities in position space are smoothed out. Instead we obtain regions of above average intensity near the positions of the would-be singularities, and regions of below average intensity elsewhere, as indicated in Fig.~\ref{figsmooth}. The typical contrast between high and low intensity regions is determined by the dimensionless {\it freak index}~\cite{hellerfreak},
\begin{equation}
\gamma =  {\delta \theta \over \Delta \theta} \approx {\delta k_x \over \Delta k_x} \sim {(u_0/v)^{2/3} \over \Delta \theta}   \,.
\label{gammadef}
\end{equation}
For example, when $\gamma \ll 1$, scattering produces small fluctuations of order $\delta k_x$ in the vertical thickness of the phase space region representing the evolved sea state in Fig.~\ref{figsmooth} (Left). The mean vertical thickness of this phase space region remains $\Delta k_x$, and the contrast ratio between maximum and average intensity therefore scales as $1+O(\gamma)$. In the opposite limit, $\gamma \gg 1$, the singularities in the ray density are only slightly softened by the spread $\Delta k_x$ in the initial sea state, and the maximum intensity is enhanced by a factor scaling as $\gamma^{3/2}$ compared to the average intensity~\cite{largegamma}.

Thus, the likelihood of freak wave formation is greatest for small initial directional spread $\Delta k_x$ and large deflections $\delta k_x$, i.e., for a well-collimated (long-crested) sea encountering a very strong random current field. On the other hand, we will see below that typical parameter values occurring in nature correspond rather to small or moderate values of the freak index, $\gamma < 1$ or $\gamma \sim 1$, where the small-$\gamma$ expansion will be more relevant. Interestingly, even in this regime where most refraction effects have been washed out, one can nevertheless observe dramatic consequences for the probability of freak wave formation, particularly for the extreme freak waves.

What justifies our focus on intensity fluctuations associated with the {\it first} generation of singularities, those formed after typical travel distance $L$? Subsequent evolution through the random eddy field causes the direction of travel to undergo diffusion, so that the initial angular spread $\Delta \theta$ becomes 
\begin{equation}
(\Delta \theta(y))^2 \approx (\Delta \theta)^2+O((\delta \theta)^2 y/L)
\end{equation}
after distance $y$, and thus the effective freak index decreases as
\begin{equation}
\gamma(y) \sim \sqrt{L \over y}
\end{equation}
at large distances $y$. The first generation of singularities, formed soon after entering a region of strong random currents, is responsible for producing the largest intensity fluctuations and thus appears to be the most hospitable environment for freak wave formation.

We now consider the implications of these spatial variations in ray density for the distribution of crest heights. Since the ray density fluctuations occur on a scale larger than a wavelength, the local wave dynamics may still be described by a Gaussian-random Longuet-Higgins model, and (\ref{rayleigh}) generalizes to
\begin{equation}
P_{\rm local}(H)=e^{-H^2/2 \tilde\sigma^2(x,y)} \,.
\label{locrayleigh}
\end{equation}
Here $\tilde \sigma^2(x,y)=\sigma^2 I_{\rm ray}(x,y)$, where $I_{\rm ray}(x,y)$ is the local density of rays,  normalized to unity for the initial sea state before scattering, and $\sigma^2$ is the elevation variance associated with the initial sea state.

Here we should note that both wave energy density and wave action density are proportional to the mean squared wave amplitude, which in turn corresponds to ray density. Thus the lumps we find in ray path density or squared wave amplitude imply, and can be converted into, lumps in either energy density or in action density. The conserved quantity in the presence of currents is wave action, rather than wave energy~\cite{garrett}. In the present circumstances, the conversion factor relating action and energy is nearly constant: the effect of currents of strength 0.5 m/sec as against a typical 10 m/sec wave velocity gives corrections on the order of 5\%.

Now let ${\cal R}_\alpha$ be the ratio of the local probability of an $H=\alpha \sigma$ event to the probability of the same event using a Rayleigh distribution. Then 
\begin{equation}
 {\cal R}_\alpha(x,y)  ={P_{\rm local}(H) \over P_{\rm Rayleigh}(H)}
 =\exp \left [{{-\alpha^2\over 2}\left ( {1\over I(x,y)} -1\right )}\right ] \,.
 \label{prob}
 \end{equation}
Eq.~(\ref{prob}) shows that local enhancement and suppression of the freak wave formation probability is very significant: in a zone where the local energy is just $50\%$ above the mean ($I=1.5$), the frequency of a $4.4 \sigma$ wave crest (the threshold for a rogue wave) increases by a factor of 25, while the frequency of a $6\sigma$ wave crest is enhanced by a factor of 400.   At the same time, the low energy zones are remarkably quiescent: $4.4 \sigma$ events are 20 times less likely in a patch down only 25\% in energy density from the mean ($I=0.75$).

After spatial averaging, we obtain the total probability of a crest height exceeding $H$,
\begin{equation}
P(H)= \int dI \, {\cal P}(I) \,e^{-H^2/2\sigma^2 I} \,,
\label{pconvol}
\end{equation}
where ${\cal P}(I) \sim \int dx\,dy \, \delta(I-I_{\rm ray}(x,y))$ is the probability distribution of energy densities obtained using ray dynamics. This local rescaling of the random wave model is unproblematic for linear wave equations, and has been used successfully in computing wave function statistics for the linear Schr\"odinger equation~\cite{lkbranch,scar} (see also Sec.~\ref{secschr}). Caution must be used in applying such rescaling to a nonlinear wave equation, as we know that over sufficiently long time scales, nonlinear instabilities will cause the sea state to re-equilibrate to a longer or shorter mean wavelength after a change in the energy density. This and related issues are addressed in Sec.~\ref{secconc}.

\subsection{Analytical Results}
\label{secanal}

We work in the regime of small or moderate freak index $\gamma$, where scattering effects are relatively weak compared with the angular spread of the incoming waves. This limit is most likely to be encountered in nature, and yet yields surprisingly large effects in the tail of the freak wave probability distribution. Consider a Gaussian-distributed local energy density, with a standard deviation $\epsilon$ proportional to $\gamma$:
\begin{equation}
{\cal P}(I) = {e^{-(I-1)^2/2 \epsilon^2} \over \sqrt{2\pi\epsilon^2}} \,,
\label{gaussianI}
\end{equation}
where
\begin{equation}
\epsilon = a \gamma\,.
\label{epsgamma}
\end{equation}
Then the distribution of crest heights becomes
\begin{equation}
P(H)= \int dI \, {e^{-(I-1)^2/2 \epsilon^2} e^{-H^2/2\sigma^2 I} \over \sqrt{2\pi\epsilon^2}} \,.
\end{equation}
For small fluctuations $\epsilon$, or large heights $H$, the integral may be evaluated by the method of steepest descent, expanding around the maximum $I-1=z$, and we obtain
\begin{equation}
P(H)=\sqrt{1+z\over 1+3z} e^{-z(1+3z/2)/\epsilon^2} \,,
\label{tail}
\end{equation}
where $z$ is defined implicitly as a function of $H$ through the cubic equation
\begin{equation}
z(1+z)^2={\epsilon^2 H^2 \over 2 \sigma^2} \,.
\label{cubic}
\end{equation}

The validity of the steepest descent approximation requires either small density fluctuations $\epsilon \ll 1$ or very tall waves $H/\sigma \gg 1$, or both, but the right hand side of (\ref{cubic}) may in general be of order unity. However, if we consider the probability of waves of a fixed height $H$ in the limiting case of a very small freak index, $\epsilon \to 0$, then $z$ becomes small and we obtain the perturbative result,
\begin{equation}
P_{\rm pert}(H)=\left[1+2 \epsilon^2{H^2 \over 4 \sigma^2}\left({H^2 \over 4 \sigma^2}-1\right)\right ]
e^{-H^2/2\sigma^2} +O(\epsilon^4)
\label{perturb}
\end{equation}
i.e., a polynomial multiplying the original Rayleigh distribution. This perturbative result is analogous to quantum wave function intensity distributions in the presence of weak disorder or weak scarring by periodic orbits~\cite{mirlin,damborsky}. According to the perturbative expression, the probability of seeing a wave height equal to or greater than $2H=4 \sigma$ (the significant wave height of the initial sea state), is unaffected by refraction, but the probability is enhanced as we consider more and more extreme waves. We note that $P_{\rm pert}(H)$ is a {\it cumulative} distribution function. The probability {\it density} of encountering a wave crest of height $H$ is given by $d P_{\rm pert}(H)/dH$, which is enhanced both for very small heights ($H<1.17\, \sigma$) and very large heights ($H>2.22\,\sigma$) and is reduced for waves of intermediate height. This is consistent with the physics of energy focusing, which increases fluctuations while keeping the average energy density unchanged.

On the other hand, if we consider the extreme tail $H/\sigma \to \infty$ for a given freak index, the right hand side of (\ref{cubic}) becomes large,  and so does the $z$ parameter. We obtain
\begin{equation}
P_{\rm tail}(H) \sim \exp{\left[-{1 \over \epsilon^2}\left[{3 \over 2} \left({\epsilon^2 H^2 \over 2 \sigma^2}\right)^{2/3}
-\left({\epsilon^2 H^2 \over 2 \sigma^2}\right)^{1/3} +{1 \over 3}+ \ldots\right]\right]} \,.
\end{equation}
Note that the leading term in the exponent grows only as $H^{4/3}$ for very large crest heights $H$, in contrast with the much faster $H^2$ growth for the Rayleigh distribution. However, depending on the value of the freak index $\gamma$ (or equivalently on the typical intensity fluctuation parameter $\epsilon$), the probabilities associated with this asymptotic tail may be too small to be observable in practice.

Also of interest are the moments of the crest height distribution, $\overline{H^{2n}}$. Since the crest heights are locally Rayleigh-distributed, we may write
\begin{equation}
H^2 = H_{\rm Rayleigh}^2 I_{\rm ray}(x,y)
\end{equation}
where $H_{\rm Rayleigh}^2$ is a Rayleigh-distributed random variable corresponding to the average surface elevation variance $\sigma^2$, and $I_{\rm ray}(x,y)$ is a slowly changing scaling factor that describes energy density variations. Averaging over space,
\begin{eqnarray}
\overline{H^{2n}} &=& \overline{H_{\rm Rayleigh}^{2n}} \; \cdot \; \overline { I^{n} } \nonumber \\
&=& \overline{H_{\rm Rayleigh}^{2n}} \; \cdot \; \int dI \, {\cal P}(I) \, I^{n} \,,
\end{eqnarray}
where the first factor is the corresponding moment in the absence of refraction, while the second factor depends only on the ray dynamics and takes into account fluctuations in the local energy density. Assuming once again a Gaussian distribution of ray densities for small freak index (\ref{gaussianI}), we have
\begin{equation}
\overline{(I-1)^{n}}=\epsilon^{n}\,(n-1)!! 
\label{gaussmom}
\end{equation}
 for even $n$ and $0$ for odd $n$, and in particular the scaling of the even moments with the freak index $\gamma$ is predicted to be
\begin{equation}
R_n = \left (\overline{(I-1)^{n}}\right)^{1 \over n} \approx \sqrt{n \over 2} \epsilon =a \sqrt{n \over 2}  \gamma.
\label{norms}
\end{equation}
In a regime where the Gaussian approximation is inadequate, the moments $\overline { I^{n} }$ or $R_n^n=\overline { (I-1)^{n} }$ may be evaluated directly from a numerical ray dynamics simulation. We note that $\overline{I}=1$ is required by probability conservation in the ray dynamics, while the higher moments $\overline { I^{n} }$ for $n\ge 2$ are necessarily greater than $1$ in the presence of refraction. However, for small freak index $\gamma \ll 1$, the corrections to Rayleigh are tiny except for the very high moments, $n \sim 1/\gamma$ for $\overline { I^{n} }$ or $n \sim 1/\gamma^2$ for $R_n$. Only in these high-order moments can one clearly see the dramatic refraction-induced effects which dominate the tail of the crest height distribution.

Finally, we must consider the effect of a finite range of wave frequencies (or speeds or wavelengths) in the initial sea state, as given for example by the JONSWAP spectrum~\cite{jonswap}. For simplicity of presentation, consider first the simplified case where the energy is uniformly distributed among all wave speeds in the interval $[v-\Delta v,v+\Delta v]$. The total finite-bandwidth ray intensity is then given by
\begin{equation}
I_{\rm fb}(x,y)= {1 \over 2 \Delta v} \int_{v-\Delta v}^{v+\Delta v} dv' \, I_{v'}(x,y) \,,
\end{equation}
where $I_{v'}(x,y)$ is the ray density associated with the single wave speed $v'$, and obeys the statistical properties discussed above for freak index $\gamma \sim (u_0/v')^{2/3}$. We are of course interested in fluctuations of the {\it total} ray intensity around its average, e.g., the variance
\begin{equation}
\epsilon_{\rm fb}^2  = {1 \over (2 \Delta v)^2} \int_{v-\Delta v}^{v+\Delta v} dv' \int_{v-\Delta v}^{v+\Delta v} dv''  \,
\overline {\delta I_{v'}(x,y) \delta I_{v''}(x,y)  } \,,
\label{vspreadeps}
\end{equation}
where
\begin{equation}
\delta I_{v'}(x,y) =I_{v'}(x,y) - 1 \sim \gamma \sim v'^{-2/3} \,.
\end{equation}
Two effects must be taken into account in determining the effect of finite but small $\Delta v$ in (\ref{vspreadeps}). First, the size of the intensity fluctuations at a given velocity is velocity-dependent, i.e.,
$\overline {\delta I_{v'}(x,y) \delta I_{v'}(x,y)  } \sim v'^{-4/3}$, and since the function $v'^{-4/3}$ has a positive second derivative, integrating over a symmetric interval of size $\Delta v$ around the central velocity $v$ will lead to an $O((\Delta v/v)^2)$ enhancement in the variance (\ref{vspreadeps}) compared with the result for a single wave velocity $v$. Secondly, replacing the initial velocity $v$ with $v'$ causes the rays to deviate by a typical distance $\sim {(v'-v) \over v} \xi$ by the time the first caustic is reached, and thus the intensity fluctuation map $\delta I_{v'}(x,y)$ is shifted
with respect to the original map $\delta I_v(x,y)$ by a similar displacement. Now assuming that the original ray intensity map  $\delta I_v(x,y)$
is sufficiently smooth on the eddy scale $\xi$ (true for large angular spread $\Delta \theta$ or small freak index $\gamma$, as discussed above), we find that the correlation between $\delta I_v(x,y)$ and $\delta I_{v'}(x,y)$ should fall off
as $1-O((v'-v)^2 /v^2)$. Both effects are second order in $\Delta v$, and combining them we find that the variance of the energy density for finite bandwidth behaves as
\begin{equation}
\epsilon^2_{\rm fb} = \epsilon^2 [1 \pm O(\Delta v / v)^2] \,,
\label{epsband}
\end{equation}
where $\epsilon$ is the zero-bandwidth result evaluated using the central velocity $v$, where the sign and numerical coefficient of the leading correction may be $\gamma$-dependent. Replacing the uniform velocity spread with a more realistic spectrum, such as JONSWAP, will merely change the numerical coefficient. In Sec.~\ref{secnumer} we will confirm that the quantitative size of the finite-bandwidth effect is very small in practice, and will almost certainly be overwhelmed by nonlinear corrections that are explicitly excluded from our model. Thus, within the linear approximation one is generally well justified in ignoring finite-bandwidth effects, and simply relying on the freak index obtained from the mean wave speed.

\section{Numerical Simulations}
\label{secnumer}
Following the work of White and Fornberg~\cite{wf}, for our numerical simulations we generate incompressible random current fields $\vec U(\vec r)$ as
\begin{equation}
U_x(\vec r) = -{\partial \psi(\vec r)}/{\partial y}; \;\;\;\;\;\; U_y(\vec r) = {\partial \psi(\vec r)}/{\partial x}\,.
\end{equation}
Here the two-dimensional stream function $\psi(\vec r)$ is Gaussian distributed with Gaussian decay of spatial correlations:
\begin{equation}
\overline{\psi(\vec r)} = 0; \;\;\;\;\;\;  \overline{\psi(\vec r)\,\psi(\vec r')} \sim e^{-(\vec r-\vec r')^2/2\xi^2}\,,
\end{equation}
and normalized so that $\overline{|\vec U(\vec r)|^2}=u_0^2$. The choice of a Gaussian-distributed and Gaussian-correlated stream function is made for convenience and for consistency with Ref.~\cite{wf}. The theoretical discussion in the previous sections does not depend on any specific choice of a random ensemble, but only on the length scale $\xi$ and velocity scale $u_0$.

\begin{figure}[htbp]
\centerline{\includegraphics[width=3.8in,angle=0]{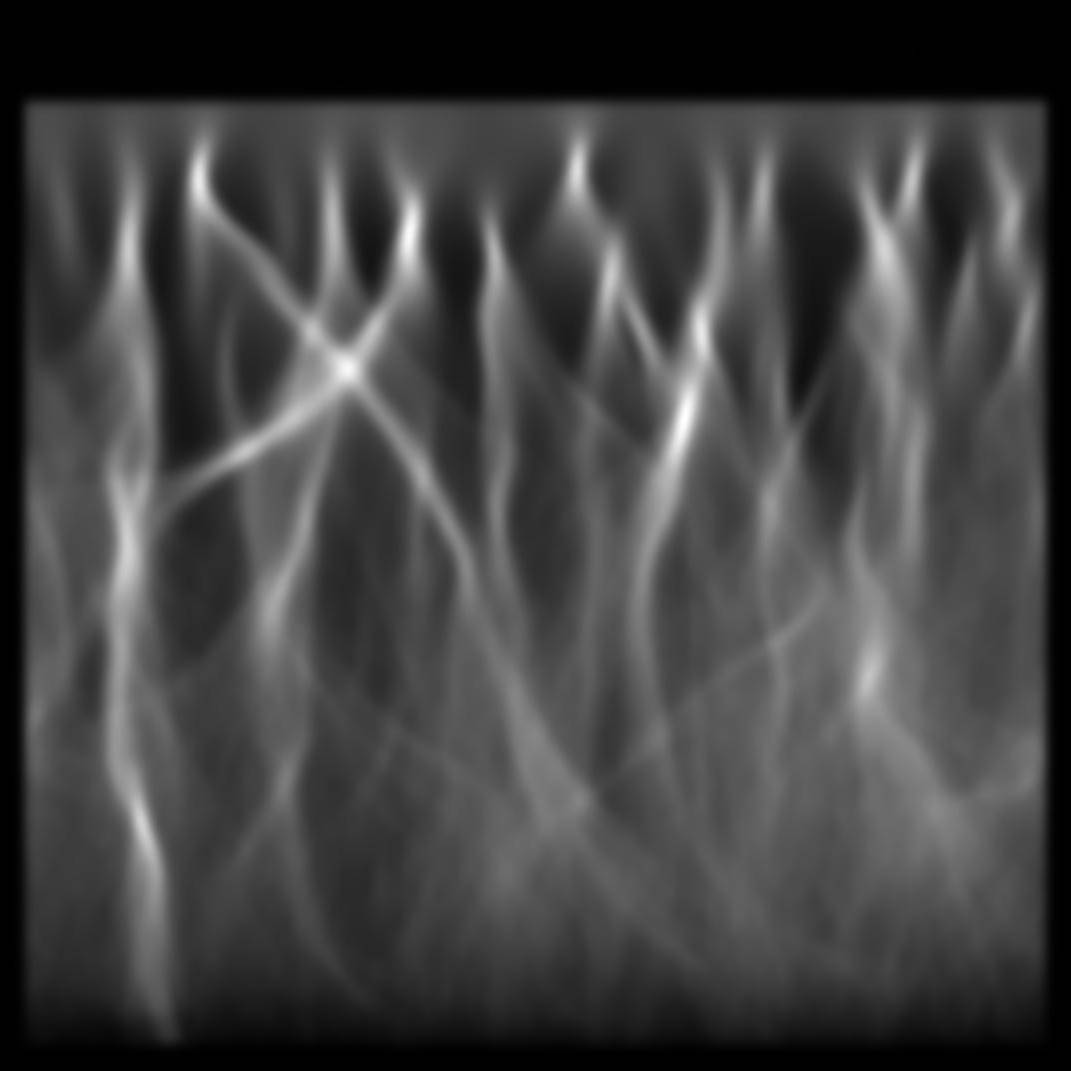}}
\vskip 0.2in
\centerline{\includegraphics[width=3.8in,angle=0]{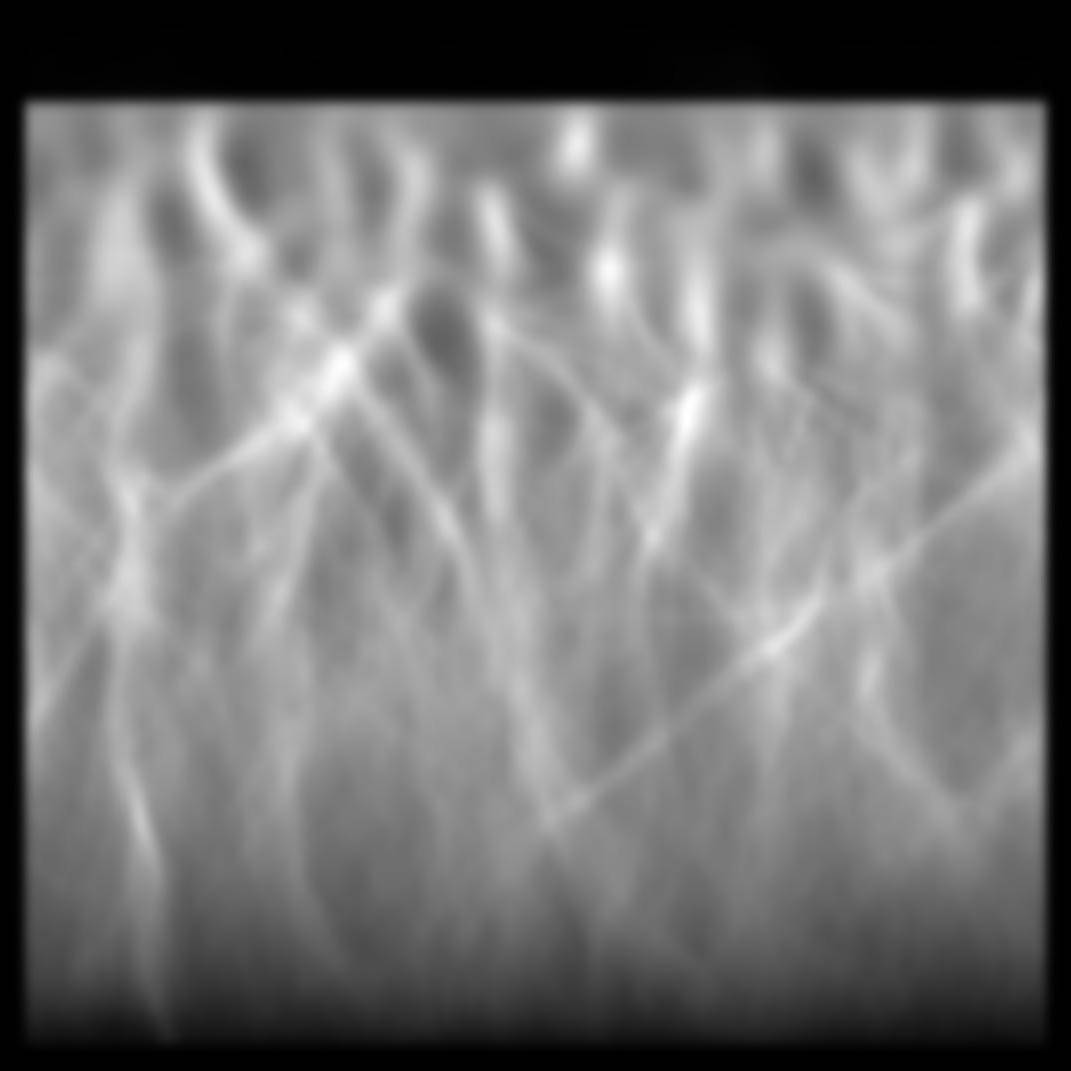}}
\caption{A ray density map is shown for rays moving through a 640 km by 640 km random eddy field with rms current $u_0=0.5$~m/s. The rays are coming in from the top with wave velocity $v=7.81$~m/s, uniform initial distribution in the $x$ direction, and initial angular spread $\Delta \theta$. Top: $\Delta \theta= 5^\circ$, corresponding to a very high freak index $\gamma=3.6$. Bottom: $\Delta \theta= 25^\circ$, or $\gamma=0.7$.}
\label{figimage}
\end{figure}

The calculation is performed on a 640 km by 640 km grid, with an eddy correlation length $\xi=20$~km and periodic boundary conditions in the transverse ($x$) direction. Without loss of generality, the rms current speed $u_0$ is set to $0.5$~m/s. Again, the specific values of $\xi$ and $u_0$ are arbitrary and serve merely to set the scale for the simulation. The refraction strength as measured by $\delta \theta$ or $\delta k_x$ may be controlled by varying the incoming wave velocity $v$, in accordance with (\ref{delkx}), while the angular spread $\Delta \theta$ is controlled directly in the initial conditions. To avoid boundary effects, ray trajectories are launched at a distance $y_0=50$~km inside the random eddy field, uniformly spaced in the transverse $x$ direction, and with a range of wave vector directions $\theta_0$. The initial wave vector for each ray is $\vec k_0=(k_0 \sin \theta_0,k_0 \cos \theta_0)$, where the initial wave number $k_0$ is chosen to correspond to constant frequency $\omega = 2\pi/(10$~sec) in accordance with (\ref{dispers}). These trajectories are interpolated and weighted with $P(\theta_0) \sim e^{-\theta_0^2/2(\Delta \theta)^2}$ to produce a ray density map $I_{\rm ray}(x,y)$. Typical density maps for the same random eddy field but two different values of the initial angular spread are shown in Fig.~\ref{figimage}.

\begin{figure}[ht]
\centerline{\includegraphics[width=3.8in,angle=0]{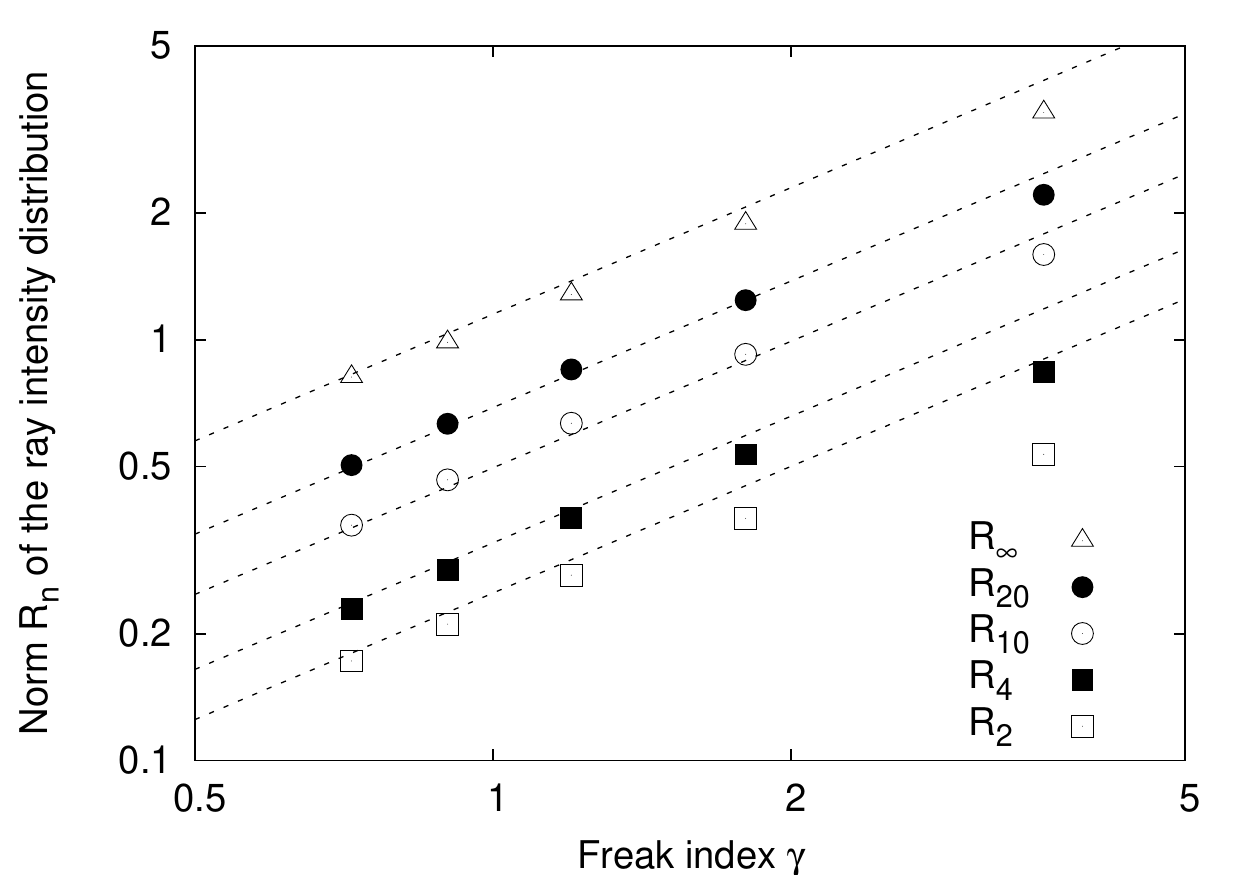}}
\caption{The norm $R_n=\left(\overline{(I-1)^{n}}\right)^{1 \over n}$ of the ray density $I_{\rm ray}(x,y)$ (Fig.~\ref{figimage}) is calculated for several $n$ and for different initial angular spreads $\Delta \theta$ (corresponding to different values of the freak index $\gamma$). From left to right, the data points represent $\Delta \theta =25^\circ$, $20^\circ$, $15^\circ$, $10^\circ$, and $5^\circ$. In each case, the intensities are sampled on a uniform rectangular grid extending in the longitudinal $y$ direction from a distance of $12.5$~km to a distance of $250$~km beyond the $y-$position where the rays are initially launched. The theoretical lines for $R_2$, $R_4$, $R_{10}$, and $R_{20}$ are obtained from (\ref{epsgamma}),~(\ref{gaussmom}).}
\label{figmom}
\end{figure}

Recalling that the median distance at which initially parallel rays form a caustic is approximately $150$~km for our parameters, we collect statistics on ray density $I_{\rm ray}(x,y)$ over the region $y_0 + 12.5$~km $<y<y_0+250$~km, which includes the most significant density fluctuations. Several moments of the ray density distribution for various initial angular spreads $\Delta \theta$ are shown in Fig.~\ref{figmom}. The initial wave velocity $v$ and rms current speed $u_0$ are kept fixed in these simulations, so varying $\Delta \theta$ corresponds to a variation of the freak index $\gamma$ (\ref{gammadef}).
\begin{figure}[htbp]
\centerline{\includegraphics[width=3.8in,angle=0]{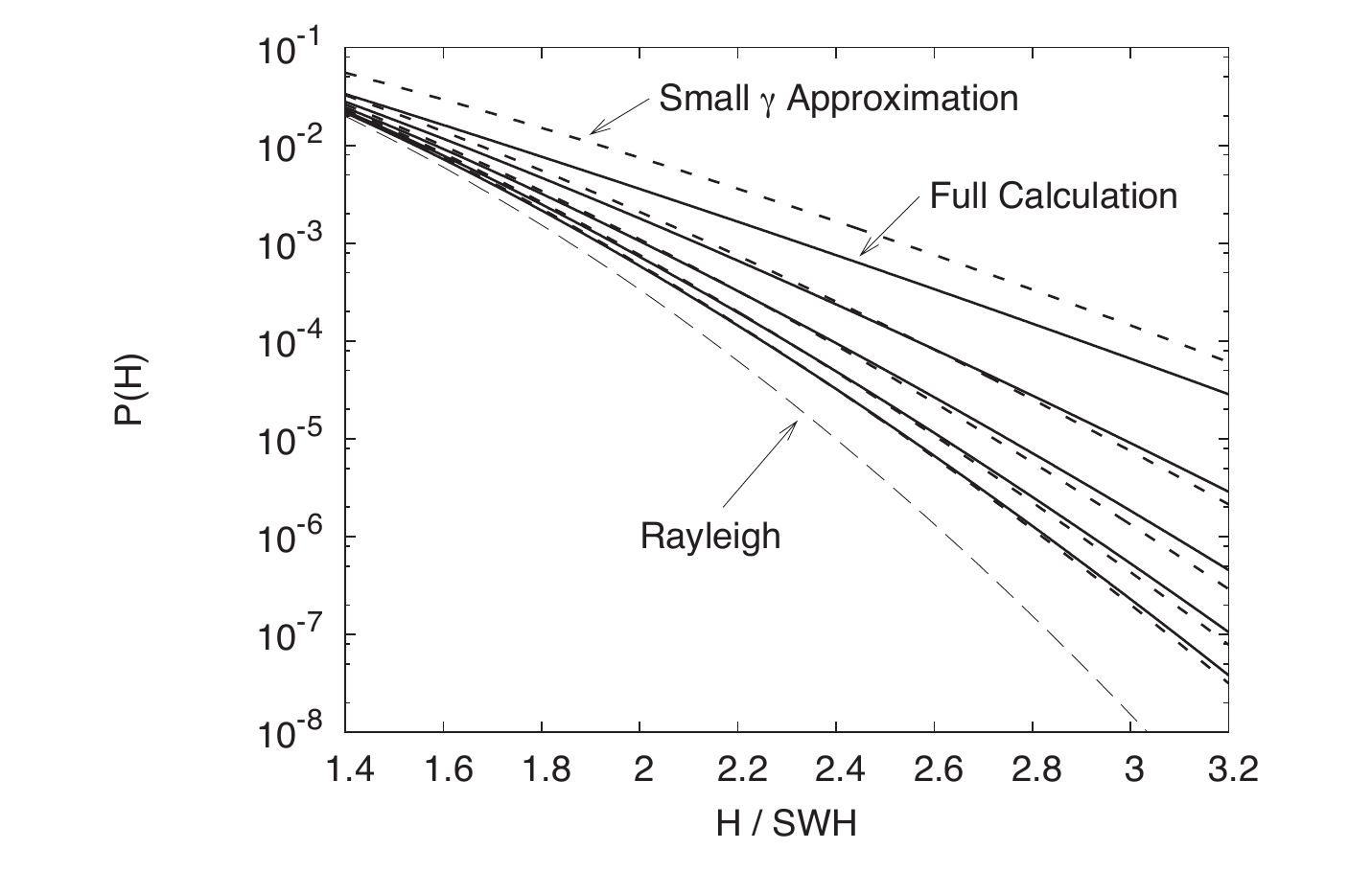}}
\caption{The fraction $P(H)$ of crest heights greater $H$ is computed from the ray densities $I_{\rm ray}(x,y)$ using the assumption (\ref{pconvol}) of locally Rayleigh fluctuations (solid lines). From bottom to top, the five solid lines show results for initial angular spread $\Delta \theta= 25^\circ$, $20^\circ$, $15^\circ$, $10^\circ$, and $5^\circ$. The small-$\gamma$ analytic prediction (\ref{tail}) is shown by a dashed line in each of the five cases. The Rayleigh distribution (\ref{rayleigh}), which describes the refraction-free $\gamma \to 0$ limit, is also shown for comparison.}
\label{figdistr}
\end{figure}

\begin{figure}[htbp]
\centerline{\includegraphics[width=3.8in,angle=0]{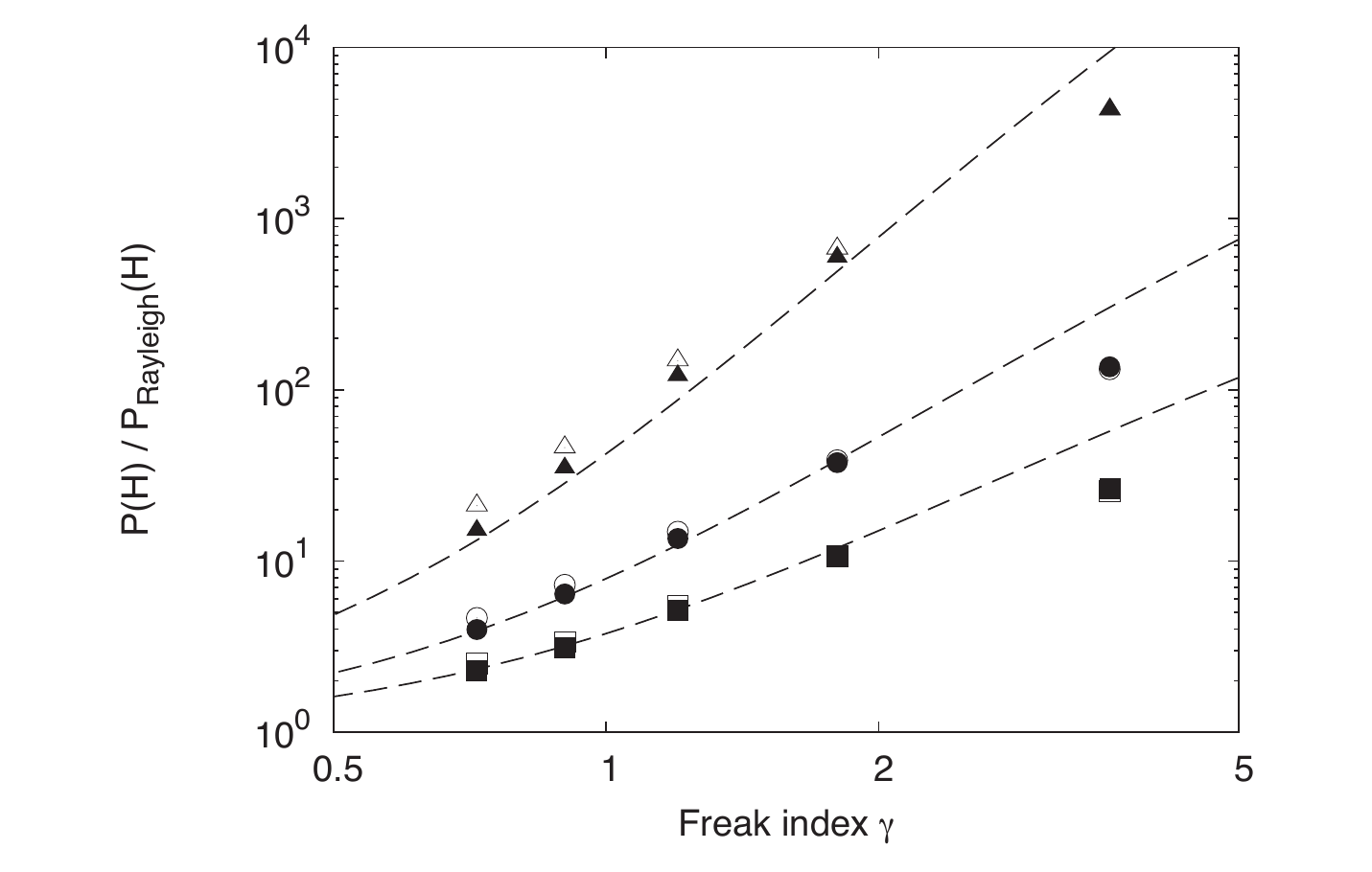}}
\caption{The increase in the number of waves with crest height greater than $H$ is shown as a function of freak index $\gamma$. Solid squares correspond to $H=4.4\sigma$ (the traditional definition of freak waves), solid circles correspond to $H=5\sigma$, and solid triangles correspond to $H=6\sigma$ (extreme freak waves). The open symbols indicate an analogous calculation, but for finite bandwidth $\Delta v/v=0.4$. From left to right, the data points represent initial angular spread $\Delta \theta =25^\circ$, $20^\circ$, $15^\circ$, $10^\circ$, and $5^\circ$. The dashed lines are obtained from the small-$\gamma$ analytic prediction (\ref{tail}).}
\label{figenh}
\end{figure}
We notice first that the norm $R_2$ is simply the standard deviation $\epsilon$ of the ray intensities, and its linear scaling with the freak index $\gamma$ for small $\gamma$ confirms the prediction (\ref{epsgamma}). Similarly the scaling of the higher moments with $\gamma$ is consistent with the prediction of (\ref{gaussmom}) and (\ref{norms}). The choice $a=0.25$ for the proportionality constant in (\ref{epsgamma}) results in good agreement with (\ref{norms}) for all moments in the regime of small or moderate freak index, as indicated by the theoretical lines for $R_2$, $R_4$, $R_{10}$, and $R_{20}$ in Fig.~\ref{figmom}. The fact that all moments are described by a single constant $a$ validates the assumption of Gaussian density fluctuations in this regime. Notably, linear dependence on the freak index $\gamma$, especially for the high moments that govern the tail of the crest height distribution, is evident for all but the largest $\gamma$ (corresponding to initial angular spread $\Delta \theta = 5^\circ$). Thus, the small-$\gamma$ approximation appears to be justified for most sea conditions expected in nature. Furthermore, the uppermost line in Fig.~\ref{figmom} shows that the maximum intensity fluctuation $R_\infty=\max(|I-1|)$ also scales linearly with $\gamma$, as expected for a finite sample.

Given the ray densities $I_{\rm ray}(x,y)$, the probability $P(H)$ of encountering a crest height larger than $H$ may be obtained using (\ref{pconvol}). The results for several values of the initial angular spread $\Delta \theta$ are shown in Fig.~\ref{figdistr}, and compared with the analytical prediction (\ref{tail}), which was derived in the limit of very small freak index. The Rayleigh prediction (\ref{rayleigh}) serves as a baseline for comparison. In Fig.~\ref{figenh}, we calculate the factor by which $P(H)$ is enhanced over the Rayleigh prediction, for several values of $H$. We see that under realistic conditions, refractive effects may enhance the probability of encountering a freak wave ($H = 4.4 \sigma$) by as much as an order of magnitude, while the probability of extreme freak waves ($H=6\sigma$) may be enhanced by more than two orders of magnitude, depending on $\Delta \theta$. The small-$\gamma$ analytic approximation provides a reasonable estimate of this enhancement for all but the least realistic situation of very long-crested incoming waves ($\Delta \theta=5^\circ$, or $\gamma=3.6$).

\begin{figure}[ht]
\centerline{\includegraphics[width=3.8in,angle=0]{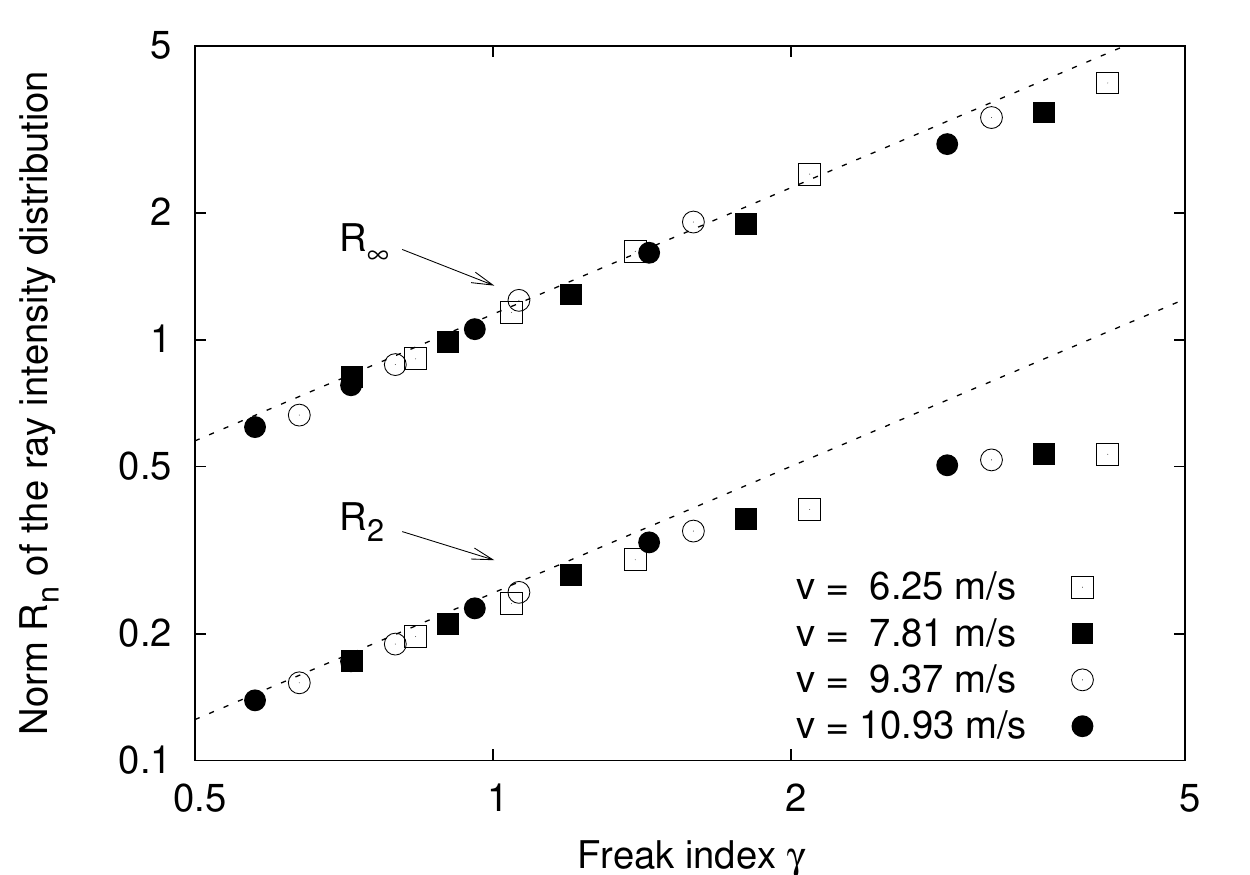}}
\vskip 0.2in
\caption{The standard deviation $R_2=\sqrt{\overline{I^2}-1}$ and maximum fluctuation $R_\infty=\max{|I-1|}$ of the ray intensity $I$ are plotted as in Fig.~\ref{figmom}, but varying the velocity $v$ of the incoming wave as well as the angular spread $\Delta \theta$ of the incoming sea. The velocities chosen correspond to $v=\beta v_0$, where $\beta=0.8$, $1.0$, $1.2$, and $1.4$, and $v_0=7.81$~m/s is the original velocity used in the previous figures. The five symbols of each type correspond to the same five values of $\Delta \theta$ as in Figs.~\ref{figmom} and \ref{figdistr}. }
\label{figvary}
\end{figure}

Until now, we have been varying the freak index $\gamma=\delta \theta/\Delta \theta$ by adjusting only the angular spread $\Delta \theta$ of the incoming sea. To verify that the freak index is in fact the single parameter determining the probability of freak wave formation in our model, we must likewise vary the refraction strength $\delta \theta \sim (u_0/v)^{2/3}$. Since the eikonal equations (\ref{eikonal}), (\ref{dispers}) are manifestly invariant under the simultaneous rescaling of the wave speed $v$ and rms current speed $u_0$ ($v \to q v$, $u_0 \to q u_0$, $k \to q^{-2}k$, $\omega \to q^{-1} \omega$), we may without loss of generality vary $v$ while keeping the currents fixed. The results of such a calculation are shown in Fig.~\ref{figvary}, and confirm that the energy density fluctuations leading to freak waves exhibit single-parameter scaling with the freak index $\gamma$.

Of course, a realistic initial sea state has a finite frequency bandwidth, and consequently includes a range of wave speeds rather than a single speed $v$. Repeating our simulations for velocity uniformly distributed in the range $[v-\Delta v,v+\Delta v]$, we find little change in the size of ray density fluctuations, as compared with the monochromatic case of a single speed $v$. This is consistent with the discussion leading to (\ref{epsband}), which showed that finite-bandwidth effects vanish at leading order in $\Delta v/v$. For example, in the typical case of freak index $\gamma=1.2$ (central velocity $v=7.81$~m/s, rms current speed $u_0=0.5$~m/s, and directional spread $\Delta \theta=15^\circ$), the moderate bandwidth $\Delta v/v=0.2$ results in an increase of less than $1\%$ in the width $\epsilon_{\rm fb}$ of the intensity distribution, as compared with the monochromatic result; doubling the bandwidth to $\Delta v/v=0.4$ still produces only a $3\%$ increase in the width of the intensity distribution. The effect on freak wave formation, for $\Delta v/v=0.4$, is indicated by the open symbols in Fig.~\ref{figenh} (the effect of bandwidth $\Delta v/v=0.2$ would be too small to be clearly visible on the scale of the figure).

Another phenomenon observed in the numerical simulations, which may be of interest when comparing with observational data, is the rapid change in mean wave direction that typically occurs in conjunction with hot spot formation. This is easy to understand in the large $\gamma$ limit, where a fold singularity results in a discontinuity in the ray density necessarily associated with a discontinuity in the mean ray direction (since the wave vector $\vec k$ associated with the fold will be different from the wave vector associated with other parts of the phase space manifold that project onto the same position $\vec r$). For finite initial angular spread $\Delta \theta$, such discontinuities are smoothed out as we have seen, but the residual effect remains. For example, on a square grid of cell size $1.25$~km ($\approx 8$ wavelengths), we observe a maximum cell-to-cell change of $25^\circ$ to $28^\circ$ in the mean wave direction when $10^\circ \le \Delta \theta \le 25^\circ$, to be compared with a median cell-to-cell fluctuation of $2^\circ$ to $3^\circ$ over a 640 km by 640 km grid for the same parameters. This rapid change in wave direction associated with entering a hot spot may be consistent with mariners' observation of some freak waves appearing at large angles from the mean wave direction.

\section{Schr\"odinger Equation Simulations and Statistics}
\label{secschr}
We now describe linear Schr\"odinger equation simulations, which are designed to check the assumptions and ideas presented so far: 1) Is the connection between mean squared wave amplitude and ray density sound, under the conditions of averaging over wavelength and especially direction?    2) is it correct to derive the global wave statistics as an integral over locally Gaussian seas (the local Rayleigh approach, see (\ref{locrayleigh}))?  
The linear Schr\"odinger equation with a random potential is a slightly imperfect laboratory for making comparisons with water waves, and of course the correct ray dynamics using eddy fields has already been presented above.  However, solving the correct nonlinear water wave equations with proper dispersion on these eddy fields  is difficult, and although we hope to do so in the future, we argue that the linear Schr\"odinger equation is adequate for the present statistical checks on ray-wave correspondence. 

In the wave simulations, statistics for rare events have to be gathered over large runs in space and time. One cannot sit only at the focal region of an eddy and have event after event; one has to wait and measure the statistics there as elsewhere.  If we had used a plane wave, the wave would repeat itself periodically. With a random superposition of hundreds of waves, we never encounter repetitions in the time allotted to the simulations. 

The simulations are performed as follows.  Both the ray and wave simulations use of course the same random potential, which plays the role of the random eddy field.  [Note that the formation of caustics and energy lumps is independent of the microscopic mechanism  of ray deflection; only the correlation length and mean deflection angle are important.]  For each run, we produce a Gaussian random potential field  $V(\vec r)$ by superposing 40 sine waves with random direction, phase, and amplitude. An overall scale factor controls the strength of the potential, and thus the distance to the first focal caustics.  The random potential may be suppressed by a smooth tanh function in the entrance region (top) and/or exit region (bottom), to simulate eddy-free zones through which the random waves travel before and after refraction by the eddies. The images shown below are about 8 correlation lengths across. 

For the ray studies, we establish a rectangular position grid. Typically 40,000 ray trajectories are launched uniformly along a line in the zero-potential entrance region, with initial direction and speed taken from appropriate Gaussian distributions.  To avoid pixelization effects, each trajectory is treated as a very narrow Gaussian density  (typically 15 pixels across, which is much less than the size of the energy lumps). 

On the same potential field we propagate waves through the region of study, using a 512 by 2048 square grid corresponding approximately to 80 by 320 wavelengths. Each incident waveset begins as an appropriate random superposition of 400-700 traveling plane waves, and is then propagated by the split operator fast Fourier transform method~\cite{split} (SOFT).
Statistics are collected at every time step.
The mean squared amplitude density plots and wave statistics data are obtained in each run (i.e., for a given potential, dispersion of incident directions, etc.) by repeating this process for 5 to 10 incident wavesets.

\begin{figure}[ht] 
   \centering
   \includegraphics[width=6in]{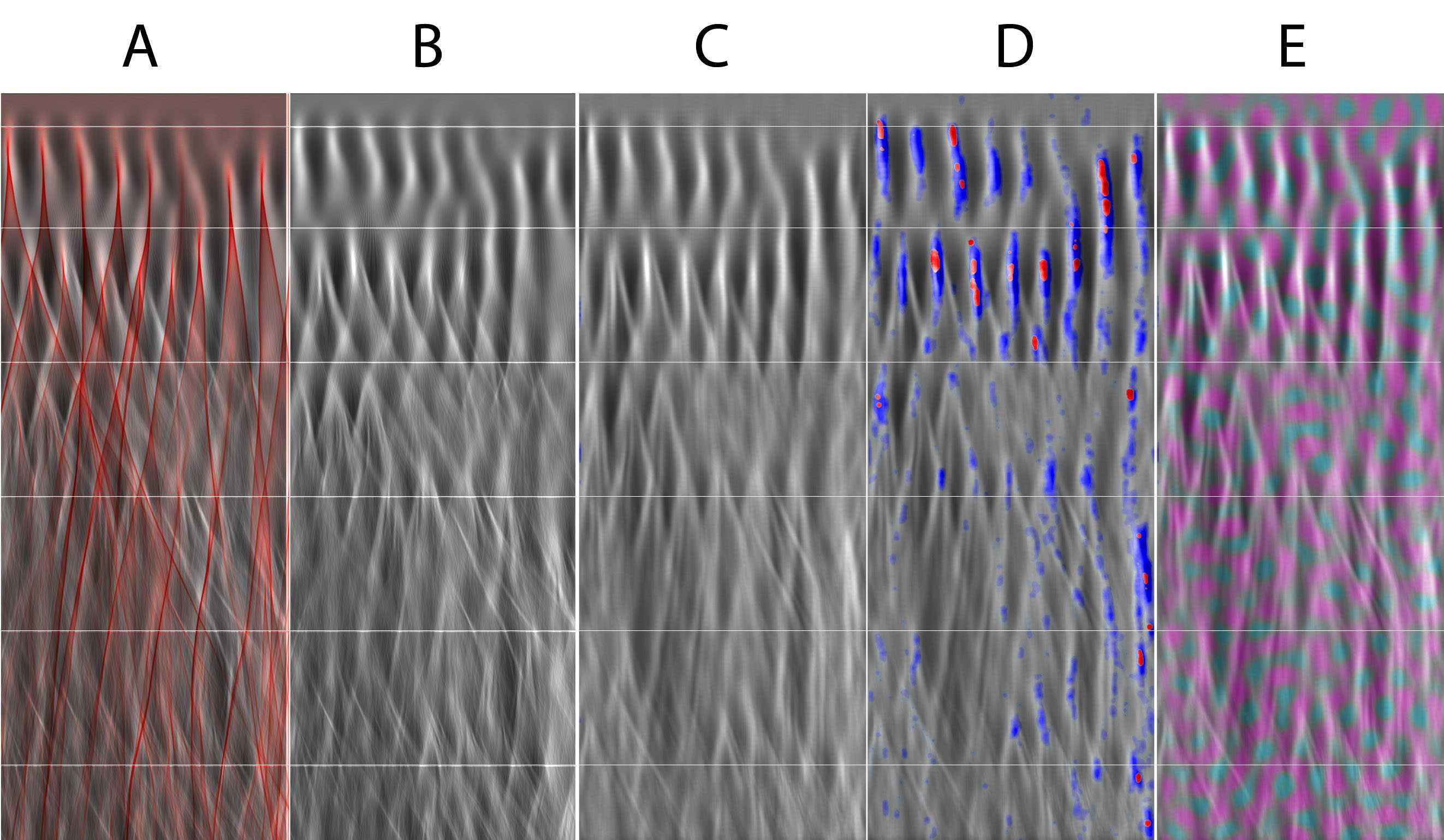} 
   \caption{Ray density (panels A, B) and wave intensity (C, D, E) data, together with the potential superimposed in E, and freak events superimposed in D, for a freak index $\gamma \approx 3.4$. Rays and waves are launched from the top toward the bottom. [Bright zones represent high average ray density or average squared wave amplitude.] {\it Very impressive and detailed agreement is seen comparing the ray density and squared wave amplitude averages.}  The classical ray density without averaging over initial direction is shown in A in red. Freak events of $4.4\sigma$ are recorded in blue in panel D, and $6\sigma$ events are shown in red. [See Fig.~\ref{freakev1} for the relevant statistical information.]
 }
   \label{set1}
\end{figure}

\begin{figure}[ht] 
   \centering
   \includegraphics[width=5in]{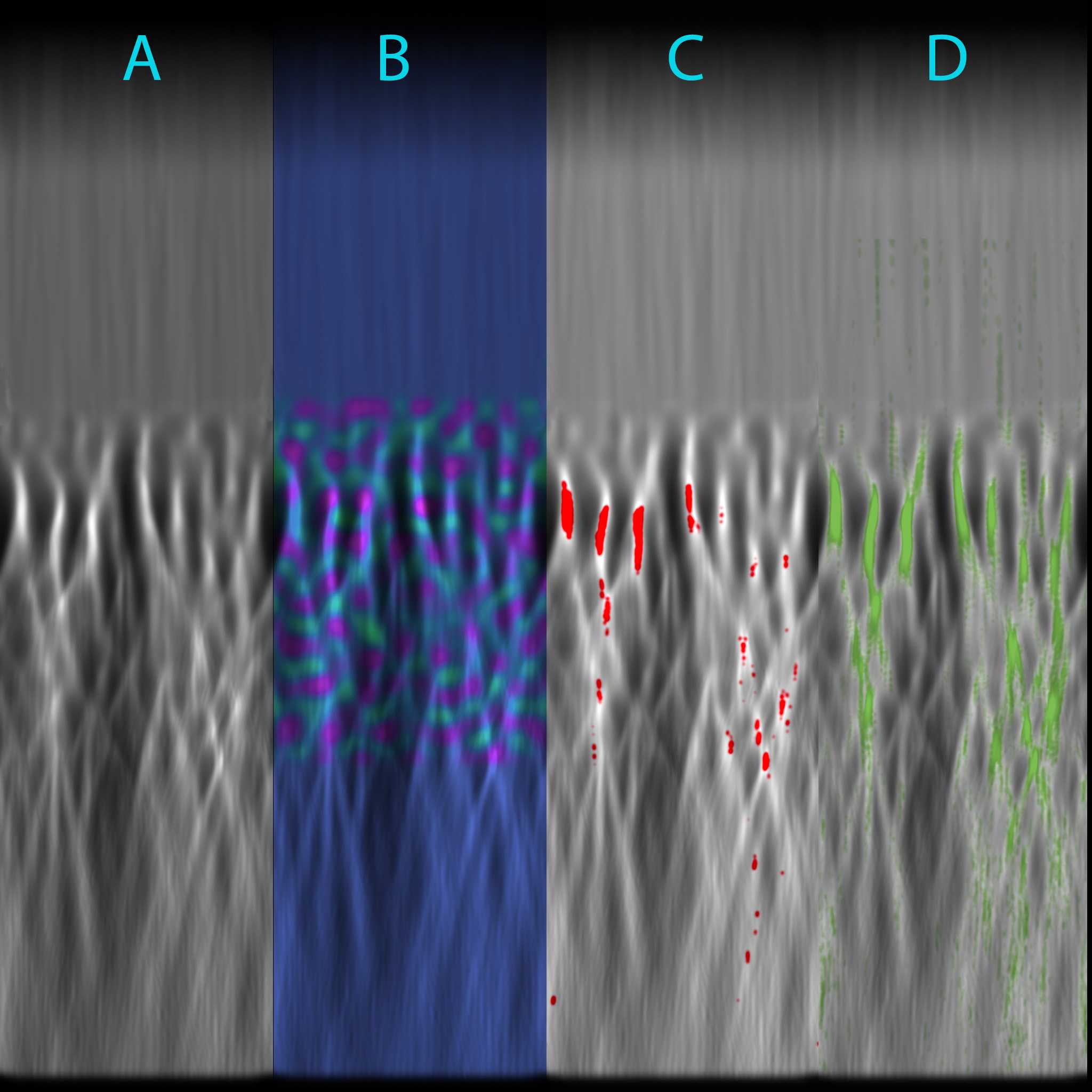} 
   \caption{All the results shown here are obtained from a wave simulation, at $\gamma \approx 2$. 700 randomly chosen plane waves, with a range of propagation directions and wavelengths, are superposed and launched from the top toward the bottom.  All four panels show the mean squared wave amplitude (bright zones indicating high mean squared amplitude), with panel A showing this quantity unadorned. Panel B superimposes the potential field, which is random only in the central region. Panels C and D superimpose freak event information, with C encoding the $6\sigma$ events, and D the $4.4\sigma$ events that occurred during the run.  }
   \label{set2}
\end{figure}

Figures \ref{set1} and \ref{set2} reveal a great deal about the calculations and the results. Since our aim is to check the accuracy of freak wave statistics predicted from ray data, all runs are performed at moderate to large freak index $\gamma$, where statistically significant numbers of freak wave events appear, and where ray-wave correspondence is most likely to be suspect. Each run takes a few hours on a single processor Macintosh G4 workstation. In Fig. \ref{set1}, we see ray density (panels A, B) and wave intensity (C, D, E) data, together with the potential superimposed in E, for a freak index $\gamma \approx 3.4$.  Very impressive and detailed agreement is seen comparing the ray density and wave intensity averages; compare B (ray) with C (wave) especially.  In A, the singular ray density for $\Delta \theta=0$ ($\gamma =\infty$, corresponding to a single incident plane wave, i.e., the White and Fornberg limit~\cite{wf}) is shown in red for comparison, superimposed on the direction-averaged ray density for $\gamma \approx 3.4$. In panel D, freak events of $4.4\sigma$ are superimposed in blue, and $6\sigma$ events (truly disastrous!) are shown in red. 

Whereas the region shown in Fig.~\ref{set1} is entirely within the random potential field, Fig.~\ref{set2} shows wave simulation results ``before, during, and after'' random scattering. Waves are again launched from the top toward the bottom. All four panels show the mean squared wave amplitude in greyscale, with panel A showing it unadorned. Panel B superimposes the potential field, which is random only in the central region and zero elsewhere. Panels C and D superimpose freak event information, with C encoding all $6\sigma$ events, and D all $4.4\sigma$ events that occurred during the run. Note that very few $4.4\sigma$ events, and no $6\sigma$ events, occur before the random potential is encountered. Significant numbers of these events are found ``downstream" of the random potential, but the largest concentration of extreme events is located within the refracting region, and especially near the first zone of (smoothed) caustics.
Referring back to (\ref{prob}), we note that the unlucky ship that finds herself in one of the bright lumps in Fig.~\ref{set2}  (where intensity is as high as four times the mean) will have approximately 1500 times the probability of encountering a $4.4 \sigma$ freak wave than in a zone of average energy density. A fearsome $5\sigma$ event is 12,000 times more likely there than  in a zone of average energy density.

     \begin{figure}[ht]
  \centerline {
 \includegraphics[width=6.in]{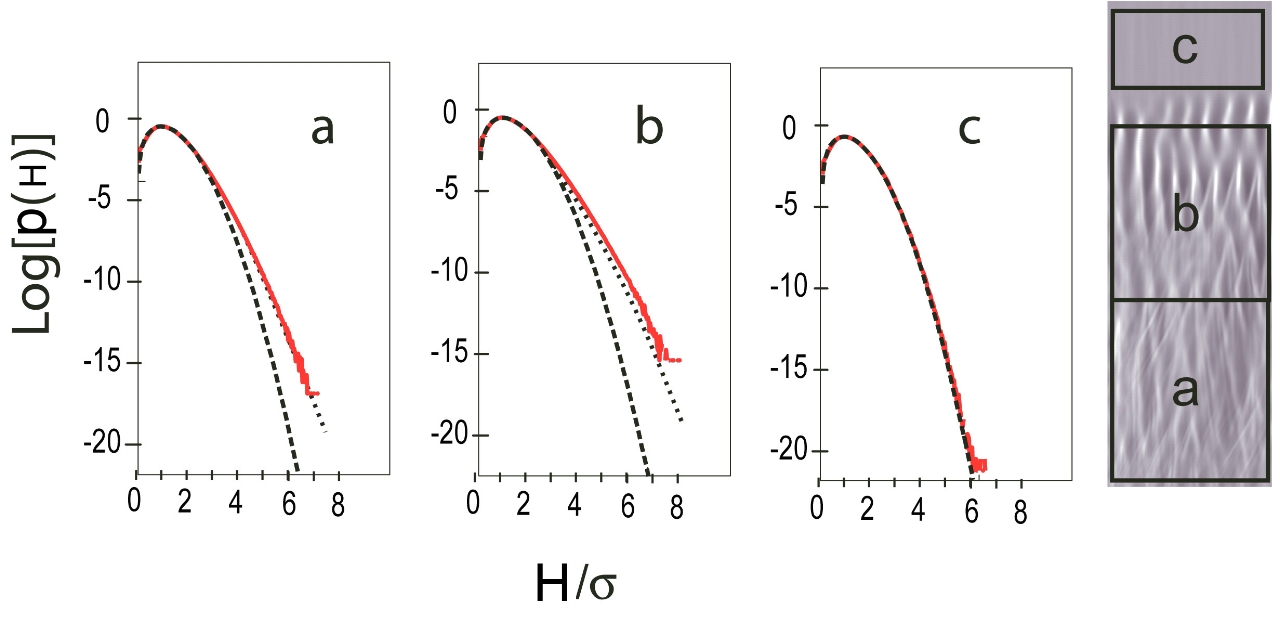}
 }
  \caption{\label{freakev1}
 Log of the crest height probability distribution by region (compared with Rayleigh and theory), for $\gamma=3.4$. The dashed line is the Rayleigh distribution, based on the average SWH. The solid line shows numerical data from wave propagation. The dotted lines represent the theory based on (\ref{pconvol}), together with measurement of the mean density $I(x,y)$. Note the excellent agreement with the Rayleigh distribution in the ``undisturbed'' region, c.}
  \end{figure}

Figure~\ref{freakev1} strongly supports our idea of locally Gaussian statistics averaged over the mean wave density distribution. The density for this run ($\gamma=3.4$) is shown in greyscale in the right panel, which also indicates the three regions c, b, a over which statistics are collected. Region c is located in the entrance zone before the onset of the random potential; region b is located inside the refraction zone and contains the first (smoothed) caustics; and finally region a shows the wave behavior after refraction.
Naturally, variations are seen from run to run and are largest for the rare events. We never obtain sufficient numbers of extreme freak wave (crest height $\ge 6\sigma$) prior to refraction (region c).  However we see many $6\sigma$ events during and after refraction.
In each case, the dashed line gives the Rayleigh distribution, based on the average SWH; the solid red line shows the numerical crest height data, and the dotted line is theory based on (\ref{pconvol}) combined with measurement of the spatial mean density variation $I(x,y)$. [That is, ${\cal P}(I)$ is determined from the mean density data $I(x,y)$, and used in the integral (\ref{pconvol}) to obtain the predicted $P(H)$.] In panel a, the energy density distribution ${\cal P}(I)$ is quite close to Gaussian. A Gaussian distribution of  energy density is also a reasonable fit to region b, which is relevant to the derivation of (\ref{tail}) in the previous Section but not needed here.  Note the excellent agreement with the Rayleigh distribution in the ``undisturbed'' region, c. Note too the excellent agreement between the data and the predicted curves in all three regions.  It is important to note that refraction effects are evident only in the tail: the probability distribution of ``typical" crest heights ($H \le 3 \sigma$) is consistent with Rayleigh in all three regions. Consequently, the measured SWH differs in the three regions a, b, and c, by less than 3\%.

\begin{table}[ht]
\begin{center}

\begin{tabular*}{0.80\textwidth}{@{\extracolsep{\fill}} | c | c | c | c | c |c |c |c |c |}
  \hline
 Run/region  & $\gamma$   &$\delta k/k$ & 6$\sigma$\ found & 6$\sigma$\  pred & 5$\sigma$\ found & 5$\sigma$\  pred & 4.4$\sigma$\ found & 4.4$\sigma$\ pred \\
    \hline
    1/c & 0.67& 0.14 & 0* & 1 & 1.4& 1& 1 & 1\\
   1/b & 0.67& 0.14 & 13 & 14 & 4 & 4& 2.5 & 2.4\\
  1/a & 0.67& 0.14 & 4.7 & 6 & 2.4 &  2.5 & 1.9 & 1.7\\
   \hline
2/c & 1& 0.22 & 0* & 1 & 0.65 & 1& 1 & 1\\
   2/b & 1& 0.22 & 41 & 47 & 7 & 9.2& 3.5 & 4.5\\
    2/a & 1& 0.22 & 10 & 4 & 3.5 &  2.5 & 2.1 & 1.7\\
      \hline
 3/c &1.35& 0.14 & 0* & 1 & 1.5& 1& 1.3 & 1\\
   3/b &1.35& 0.14 & 356 & 349 &26 &27&8 & 8\\
   3/a & 1.35& 0.14 & 148 & 106 & 15 &  13 & 6 &5\\
   \hline

\end{tabular*}
 
\end{center}
\label{runs}
\caption{ For three typical runs shown here, we report events of size at least $4.4\sigma$, $5\sigma$, and $6\sigma$ in regions a, b, and c. See Fig.~\ref{freakev1} for an image of the three zones. Runs 1 and 3 have the refracting zone extending throughout regions b and a. Run 2 has the refracting zone only in region b (as in Fig.~\ref{set2}). Region c is the non-refracting zone preceding first entry into the refraction region. Here ``$n\sigma$ pred'' is the predicted number of $n\sigma$ events ($n=6$, $5$, $4.4$), divided by the expected number based on Gaussian statistics over the whole region. Similarly ``$n\sigma$ found'' is the theoretically predicted value for this ratio. The asterisk indicates that no $6\sigma$ events were seen in the pre-refracting region for the entire run. For a freak index $\gamma=1$, we see a factor of $\sim 50$ increase in the number of $6\sigma$ and larger events, a factor of $\sim 10$ increase in $5\sigma$ and larger events, and a $\sim 4$ fold increase in $4.4\sigma$ and larger events, compared with the Longuet-Higgins Gaussian seas model in the refracting region b. Run 1 for $\gamma=0.67$ ($\Delta \theta \approx  9$ degree spread of incident wave directions, and $\delta \theta \approx 6$ degree mean deflection) shows correspondingly lower enhancement of freak events over the Gaussian expectations. Run 3, for a freak index of $1.35$, shows up to 350 fold enhancement of 6$\sigma$ and larger events.
}
\end{table}%

Table \ref{runs} presents some typical data obtained from the linear Schr\"odinger propagation. For the three typical runs shown here, we report $4.4\sigma$, $5\sigma$, and $6\sigma$ and larger events in regions a, b, and c. See  Fig.~\ref{freakev1} for an image of the three zones.

\section{Qualitative Observations}
\label{qual}

 We remark briefly on some qualitative and perhaps speculative aspects of the results. Anecdotal reports from mariners implicate at least three types of freak waves: 1) The ``three sisters'', i.e., three large waves in a row heading in approximately the mean wave direction, 2) ``out of nowhere'' waves, which attack suddenly from an unexpected quarter, perhaps 45 degrees from the mean wave direction, and 3) the infamous ``hole in the sea followed by a wall of water''.  This third kind of wave is reported to be persistent, and perhaps very broad. Certainly, nonlinear effects are important in forming and sustaining the third type of wave; it is possible that the linear effects considered here could help trigger it.

There is modest evidence for the first two types of wave in our simulations.  This is not to say that nonlinearities play no role, indeed they must for a complete description.  But we find in the simulations that the freak events often appear as three sisters, or perhaps five with the middle three being tallest, for the parameters we used.  Furthermore, we do see rays propagating at high angle to the mean direction; these leave tracks such as the small, fairly bright branches seen in Fig.~\ref{figimage}, even in the case of a 25 degree spread in the incident waveset. Such streaks are also seen in the Schr\"odinger simulations in Figs.~\ref{set1} and \ref{set2}. Waves are seen traveling in the direction of the streaks in the simulations. These streaks are typically guided by remnant fold caustics that have survived the averaging and received several chance deflections in one direction.  They are compact and move at a high angle relative to the mean wave direction. We call these streaks and the waves that populate them ``runners''.

We should also mention another mechanism for lumping, namely the collision or overlapping of two smoothed  V-shaped fold caustics that form immediately following a cusp. These resemble ``rooster tails'' seen in the wake of power boats, and we call them by that name. Several are seen in Figs.~\ref{set1} and \ref{set2}. They cause a sudden increase of local average wave action. 

Certainly, the qualitatively reasonable idea that the ray density fluctuations are washed out due to chaotic exponential instability of the rays is incorrect. What is perhaps even more surprising is that instead of smoothing out, the ray density develops finer structure after propagating several times the distance to the first focal regions.

\section{Conclusions and Outlook}
\label{secconc}

We have seen that refraction of a stochastic Gaussian sea by random eddy currents creates a lumpy spatial energy distribution, causing deviations from the initial Rayleigh crest height (or wave height) distribution expected in a random wave model. Significant energy lumps survive averaging over wave direction and wavelength, despite the chaoticity displayed by individual ray trajectories and the washing out of singularities in the ray dynamics. These lumps dramatically increase the probability of freak wave formation, even though parameters associated with low-order moments of the wave height distribution, such as the significant wave height or the kurtosis, are almost unchanged. A single dimensionless parameter called the freak index $\gamma$, defined as the ratio of typical angular deflection in one focal distance to the initial angular uncertainty of the incoming waveset, is sufficient to determine the size of the tail in the crest height distribution. Although freak wave probability increases with increasing $\gamma$, very significant effects are obtained already when $\gamma$ takes modest values realizable in nature. The increase in extreme freak waves (those of wave height greater than three times the significant wave height) is especially spectacular. The number of such waves may increase by two or more orders of magnitude when a well-collimated sea (angular spread $\Delta \theta \le 15^\circ$) encounters a field of strong eddy currents (rms current speed $u_0=0.5$~m/s). For physically reasonable parameters, good agreement is obtained between analytic results based on a small$-\gamma$ approximation and numerical simulations.

In this paper we use statistics averaged over a large area, including all the lumps and streaks, to define the SWH and the global statistics.  For any reasonable freak index, the low moments of the global distribution, including the kurtosis, are scarcely affected, yet as we have shown the far tails of the distribution, i.e., the ``freak" events, can be greatly enhanced.   One could perhaps  argue that inside a spatially extended lump one should re-define the SWH to adjust to the local energy density and then inside that region the statistics would be the ``expected'' Gaussian with a larger SWH than say some kilometers away. The largest of  the lumps are on the order of the correlation length of the eddies,  which could be thought of as either large or small depending on the wavelength of the seaway. There are also much smaller lumps that appear when two streaks cross for example, as well as the fine structure mentioned in the previous Section. The basic idea of the non-uniform sampling theory used here is that the seas suffer a {\it sudden} increase in local wave action when encountering a lump, and they have neither the time nor space to fully accommodate this increase through the normal nonlinear evolution. The seas inside a lump have the same wavelength but larger amplitude and are therefore steeper, greatly enhancing the probability of a freak wave appearing, just as in other scenarios where wave steepness increases. By combining refraction and random wave statistics we retain a statistical model for freak wave formation. 

As emphasized in the introduction, the linear model considered here may be regarded a starting point for a more sophisticated nonlinear analysis. To leading order in the wave steepness, nonlinear effects may be taken into account by replacing the Rayleigh distribution of crest heights in (\ref{locrayleigh}) with the Tayfun distribution~\cite{tayfun}. At this order, the only effect of nonlinearity is to create an asymmetry between wave crests and wave troughs, while the distribution of wave {\it heights} remains unchanged. Numerical evidence suggests that the Tayfun approximation gives accurate results for crest height statistics in two dimensions except in the case of a narrow directional spread~\cite{sdtkl}.

At higher order, nonlinearity of the wave equation results in exponential instabilities, such as the Benjamin-Feir instability~\cite{bf} for an initial plane wave evolved under the nonlinear Schr\"odinger equation (NLS). The likely effect of such instabilities is an enhancement in the probability of freak wave formation on short to moderate distance scales~\cite{nonlin1,nonlin2} (the Benjamin-Feir time scale is of order $(kH)^{-2}$ wave periods, where $kH$ is the steepness). On longer time scales, nonlinearity should have the opposite effect of reducing wave steepness by transferring energy in the hot spots to longer wavelength modes, and resulting in a new equilibrium distribution with a larger significant wave height. We do not expect such re-equilibration to be effective when the energy density is varying significantly over scales as short as $10$ wavelengths.  However, further investigation employing a nonlinear model such as a four-wave approximation~\cite{zakharov} is needed to determine whether refraction and nonlinearity acting together produce more freak waves than does either effect separately. Interestingly, recent numerical explorations of nonlinear effects on freak wave formation in two dimensions (using nonlinear equations of the Dysthe type) show that these effects are strongly dependent on directional spread~\cite{sdtkl,oos}. Nonlinear enhancement of freak wave formation is maximized for long-crested waves, which is the same limit in which refraction effects are greatest. Of course, nonlinear effects additionally scale with the initial wave steepness (e.g., with the $\alpha$ parameter in the JONSWAP spectrum), while refraction effects depend on scattering angle, allowing for nontrivial interplay between the two mechanisms. A number of other issues suggest themselves but are not addressed here, such as experimental detection of energy lumpiness, and the rate of movement of lumps (due to changing eddy positions and velocities).

The essence of this paper is a synthesis of Longuet-Higgins Gaussian seas~\cite{randseas} and the refraction model of White and Fornberg~\cite{wf}. From the perspective of wave propagation, it is still a linear theory, and indeed we have checked it successfully with a linear Schr\"odinger propagator. From the point of view of ray dynamics it is decidedly nonlinear and even chaotic in the usual sense of exponential sensitivity to small changes in initial data.  If the Longuet-Higgins Gaussian  seas model is ``too cold'' (too few extreme events are predicted to occur), and the White and Fornberg model is ``too hot'' (freak waves appear periodically at every focal point), then the present combination of Gaussian seas and refractive effects may be ``just right''.

We mention that random refraction due to many small eddies is not essential to the main idea presented here, which is the notion of random wave statistics in the presence of energy lumps and streaks.  Such variations in energy or action could have many causes, including a single large eddy or indeed a variable shallow bottom.  In fact the latter seems the best hope for a real world laboratory to test the ideas in this paper. For a sea incident on a continental shelf with bottom depth variations, one might hope to compare the statistics of the incoming wave with the statistics at various locations that are predicted to be action maxima or minima. This could hopefully become a semiquantitative test of the ideas presented here, i.e., a deviation from Gaussian statistics in the far tails of the distribution. Indeed the Canadian waters off Vancouver Island and the Queen Charlotte Islands, an area known for its unruly seas, may be one such ``laboratory''. 

Finally we express our conviction that nonlinear wave effects are very important, perhaps to every freak wave event.  However, we believe the sudden buildup of energy or action in the lumps that we have shown exist may be an important triggering mechanism for nonlinear evolution. 

\begin{acknowledgments}
This work was supported in part by the National Science Foundation under Grant No. PHY-0545390 and by Louisiana Board of Regents Support Fund Contract No. LEQSF 2004-07-RDA-29.  EJH is very  grateful for helpful suggestions and conversations with Chris Garrett, Johannes Gemmrich, and Rick Thomson.
\end{acknowledgments}


\begin{thebibliography}{99}
\bibitem{soften} The typical structure of the phase space manifold near a fold singularity is ${x-x_0\over \xi} \sim   \left( k_x -k_{x0} \over \delta k_x\right)^2$; the uncertainty principle implies that the smallest phase space area enclosed by the fold that can be resolved by wave dynamics is $(x-x_0)(k_x-k_{x0}) \sim 1$.
\bibitem{largegamma} Near a fold singularity, the manifold ${x-x_0\over \xi} \sim   \left( k_x -k_{x0} \over \delta k_x\right)^2$ is thickened by $O(\Delta k_x)$ in the $k-$direction and $O(\xi {\Delta k_x \over \delta k_x})$ in the $x-$direction.
\end{thebibliography}
\end{document}